\documentclass[conference]{IEEEtran}
\IEEEoverridecommandlockouts
\usepackage{cite}
\usepackage{amsmath,amssymb,amsfonts}
\usepackage{algorithmic}
\usepackage{graphicx}
\usepackage{textcomp}
\usepackage{xcolor}
\usepackage{multirow}
\usepackage{kotex}
\usepackage{multicol,multirow}
\usepackage{makecell}
\usepackage{booktabs}
\usepackage{stackengine}

\def\BibTeX{{\rm B\kern-.05em{\sc i\kern-.025em b}\kern-.08em
    T\kern-.1667em\lower.7ex\hbox{E}\kern-.125emX}}
\begin{document}

\title{Improving Generalization of Drowsiness State Classification by Domain-Specific Normalization\\
\thanks{This work was supported by the Institute of Information \& Communications Technology Planning \& Evaluation (IITP) grant, funded by the Korea government (MSIT) (No. 2019-0-00079, Artificial Intelligence Graduate School Program (Korea University); No. 2021-0-02068, Artificial Intelligence Innovation Hub).
}}
\makeatletter
\newcommand{\newlineauthors}{%
  \end{@IEEEauthorhalign}\hfill\mbox{}\par
  \mbox{}\hfill\begin{@IEEEauthorhalign}
}
\makeatother

\author{\IEEEauthorblockN{Dong-Young Kim}
\IEEEauthorblockA{\textit{Dept. of Artificial Intelligence}\\
    \textit{Korea University}\\
    Seoul, Republic of Korea \\
    dy\_kim@korea.ac.kr}
\and
\IEEEauthorblockN{Dong-Kyun Han}
\IEEEauthorblockA{\textit{Dept. of Brain and Cognitive Engineering} \\
    \textit{Korea University}\\
    Seoul, Republic of Korea \\
    dk\_han@korea.ac.kr}
\newlineauthors
\IEEEauthorblockN{Seo-Hyeon Park}
\IEEEauthorblockA{\textit{Dept. of Brain and Cognitive Engineering} \\
    \textit{Korea University}\\
    Seoul, Republic of Korea \\
    tjgus9190@korea.ac.kr}
\and
\IEEEauthorblockN{Geun-Deok Jang}
\IEEEauthorblockA{\textit{Dept. of Artificial Intelligence} \\
    \textit{Korea University}\\
    Seoul, Republic of Korea \\
    gd\_jang@korea.ac.kr}
\and
\IEEEauthorblockN{Seong-Whan Lee}
\IEEEauthorblockA{\textit{Dept. of Artificial Intelligence} \\
    \textit{Korea University}\\
    Seoul, Republic of Korea \\
    sw.lee@korea.ac.kr}
}

\maketitle

\begin{abstract}
Abnormal driver states, particularly have been major concerns for road safety, emphasizing the importance of accurate drowsiness detection to prevent accidents. Electroencephalogram (EEG) signals are recognized for their effectiveness in monitoring a driver's mental state by monitoring brain activities. However, the challenge lies in the requirement for prior calibration due to the variation of EEG signals among and within individuals. The necessity of calibration has made the brain-computer interface (BCI) less accessible. 
We propose a practical generalized framework for classifying driver drowsiness states to improve accessibility and convenience. We separate the normalization process for each driver, treating them as individual domains. The goal of developing a general model is similar to that of domain generalization. The framework considers the statistics of each domain separately since they vary among domains. We experimented with various normalization methods to enhance the ability to generalize across subjects, i.e. the model's generalization performance of unseen domains. The experiments showed that applying individual domain-specific normalization yielded an outstanding improvement in generalizability. Furthermore, our framework demonstrates the potential and accessibility by removing the need for calibration in BCI applications.

\end{abstract}


\section{INTRODUCTION}
Brain-computer interface (BCI) is a technology that connects our brains to external devices or computers \cite{wolpaw2000brain, thung2018conversion}. Among the different types of BCI, using electroencephalogram (EEG) signals to detect emotional and mental states is referred to as passive BCI \cite{wu2020transfer}. Passive BCI systems don't require us to actively think about things unlike active BCI \cite{lee2019towards, lee2020neural, al2021deep, lee2019comparative} and quietly observe our mental states. Affective BCI, which is a part of passive BCI, focuses on understanding and recognizing our feelings and emotions \cite{houssein2022human, kim2015abstract} like motion sickness \cite{bang2023ms}, drowsiness \cite{cui2022eeg}. These states are detected by EEG, galvanic skin response, heart rate, etc.

Abnormal driver states, including drowsiness and motion sickness, have long been a major cause of road accidents \cite{tefft2010asleep}. Therefore, early detection and prevention of abnormal drive states is necessary \cite{higgins2017asleep}. Among various monitoring methods, physiological signals, particularly electroencephalogram (EEG) signals, are effective for monitoring driver states because they directly measure brain activity, reflecting the driver's mental condition \cite{lee2019possible, suk2014predicting}. 


EEG signal, on the other hand, is a challenging method to use due to its intra-variability and inter-variability \cite{kostas2020thinker}. It varies based on factors such as measurement duration, the person's physical condition, and mood, and the placement of sensors. Typically, a calibration session is needed to fine-tune the system for each user by collecting user-specific data for about 30 minutes. Calibration is not only time-consuming but can also result in a negative user experience. Therefore, transfer learning \cite{kim2019subject}, especially domain generalization (DG) has been widely used to enhance user convenience by eliminating calibration.

We addressed the problem from a DG perspective by considering users as independent domains and treating the variability as a domain shift. The goal of developing a generalized drowsiness classification framework without calibration is similar to DG \cite{wu2020transfer}. There has been an increase in interest recently in specifying it as a DG task. Cui \textit{et al.} \cite{cui2019eeg} extracted power spectral density features and applied episodic training, a DG method in computer vision. Kim \textit{et al.} \cite{kim2022dg} applied augmentation and regularization inspired by DG methods.

In particular, normalization in deep neural networks has shown an impact on DG tasks \cite{zhou2022domain, wang2022generalizing}. A straightforward approach would involve training a deep neural network with domain-invariant normalization, applying all training samples regardless of their respective domain. However, as statistics vary across different source domains, using mixed statistics from multiple source domains can contaminate the network from learning generalizable representations \cite{seo2020learning}. Therefore, a solution has emerged in the form of domain-specific normalizations \cite{Chang_2019_CVPR} to capture domain-specific statistics. Seo \textit{et al.} \cite{seo2020learning} proposed a domain-specific optimized normalization method in which each domain has its mixture of instance normalization (IN) layer and batch normalization (BN) layer. Moreover, various methods are proposed in how to report the final results based on the output of the model. Segu \textit{et al.} \cite{segu2023batch} proposed an aggregation method that weighs the difference between the instance-level feature statistics of test data and the BN statistics of the source domain.

In summary, preventing accidents caused by drowsy driving is crucial. While EEG signals are known to be useful for estimating drowsiness, calibration can decrease the accessibility and usability of EEG-based classification models. We propose a framework that is invariant to the user or driver changes for driver drowsiness state classification. In other words, we propose a framework that demonstrates outstanding generalization performance. We experimented with multiple normalization methods and claimed that domain-specific batch normalization specifically using the average of probabilities at classification for inference enhances the ability to generalize across subjects in the EEG signal dataset.

\section{METHODS}
We conducted a comparison of three different deep neural networks in a total of seven models with various numbers of layers and kernel sizes. Subsequently, the model with the highest performance was chosen as the backbone model for experimenting with different normalization and inference methods.

\subsection{Deep Neural Networks} 
We briefly introduce three deep neural networks: DeepConvNet \cite{schirrmeister2017deep}, EEGNet4,2 \cite{lawhern2018eegnet}, EEGNet8,2 \cite{lawhern2018eegnet}, ResNet1D-8 \cite{han2021domain}, and ResNet1D-18 \cite{kim2022dg}. DeepConvNet and EEGNet are commonly utilized in the field of EEG-based classification \cite{mane2021fbcnet, bang2021spatio}, and ResNet1D has demonstrated remarkable performance in tasks related to DG \cite{han2021domain} and mental state classification \cite{kim2022dg, kim2023calibration}. Furthermore, we made adjustments to the residual blocks in ResNet1D-8 and ResNet1D-18 than in \cite{han2021domain} to enhance classification.  
\subsubsection{DeepConvNet}
DeepConvNet comprises four convolution-max-pooling blocks. In the first block, temporal convolution and spatial convolution are performed. The subsequent three blocks are standard convolution max-pooling blocks, featuring the exponential linear unit as the activation function.

\subsubsection{EEGNet}
EEGNet4,2 learns from four temporal filters and two spatial filters per temporal filter. On the other hand, EEGNet8,2 has the same structure as EEGNet4,2 but differs in learning from eight temporal filters and two spatial filters per temporal filter. Additional model-specific details can be found in \cite{kim2023calibration}.

\subsubsection{ResNet1D}
ResNet1D-8 comprises three residual blocks and a fully-connected layer. Each residual block consists of one-dimensional convolutional layers, dropout, BN, and a skip connection, which includes a convolution layer and BN layer when there is a discrepancy between the input and output size of the residual block. Similarly, ResNet1D-18 comprises four residual blocks and a fully-connected layer. Additional model-specific details can be found in \cite{kim2023calibration}. In addition, we modified the order of the layers within the residual blocks as shown in Fig. \ref{fig:resblock} and used the Gaussian error linear unit as the activation function.

\begin{figure}[t!]
    \includegraphics[width=0.8\columnwidth]{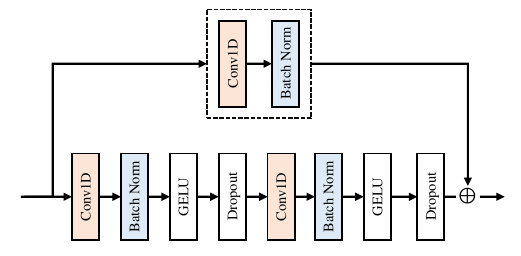}
    \centering
    \caption{Illustration of the residual block. An additional convolution layer and batch normalization layer are used for skip connection when the input and output size of the residual block differs. Gaussian error linear unit (GELU) was used as the activation function.}
\label{fig:resblock}
\end{figure}

\begin{figure}[t!]
    \includegraphics[width=\columnwidth]{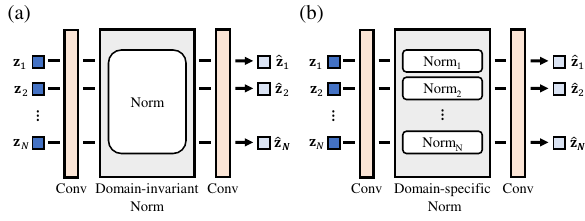}
    \centering
    \caption{Illustration of the two types of normalization in the residual block. (a) Domain-invariant normalization layer (b) Domain-specific normalization layer. $\textbf{z}$ denotes the trained intermediate features, $\widehat{\textbf{z}}$ denotes the normalized features, and $N$ denotes the number of domain.}
    \label{fig:bn}
\end{figure}

\subsection{Normalization Methods}
We experimented with two types of normalization (BN, IN) and a combination of them (instance batch normalization (IBN)). Moreover, we experimented with domain-invariant normalization, i.e. sharing one normalization layer among domains, and domain-specific normalization, i.e. having separate normalization layers across domains. Fig. \ref{fig:bn} shows the two types of normalization used for comparison.

\subsubsection{Domain-invariant normalization}
As shown in Fig. \ref{fig:bn}(a), deep neural networks are generally trained with a domain-invariant normalization using all training samples regardless of their domain. We compared two types of conventional normalization techniques and an addition of the combination of them which has recently gained interest.

\paragraph{Batch normalization}
Batch normalization (BN) \cite{ioffe2015batch} is a commonly used normalization method, which normalizes features at a mini-batch level. BN preserves the instance-level style variation but degrades performance when trained with domains that have a big difference. The batch statistics overfit to a particular domain which results in a degradation in the generalization performance in unseen domains.

\paragraph{Instance normalization}
Instance normalization (IN) \cite{ulyanov2016instance}, unlike BN which operates per mini-batch, applies the same process but on an individual instance basis. It is recognized for its efficiency in eliminating the specific characteristics of each instance. In other words, IN reduces the style information in each domain. However, compared to BN, the normalized features are less discriminative across different classes.

\paragraph{Instance batch normalization}
Instance batch normalization (IBN), a mixture of IN and BN, enhances the benefits of IN and maintains the classification performance. IBN optimizes the trade-off between maintaining differences across categories and achieving consistency across various domains \cite{pan2018two, seo2020learning}.

\subsubsection{Domain-specific normalization}
We separated the normalization method (domain-specific BN (DSBN) \cite{Chang_2019_CVPR}, domain-specific IN (DSIN), and domain-specific optimized normalization (DSON) \cite{seo2020learning}) to address domain shift and generate domain-invariant representations to improve generalizability.

\paragraph{Max logit and max probability}
Two options can be chosen as the final result, which is based on the max logit ($max(logit_1,...,logit_N)$) or max probability ($max(prob_1,...,prob_N)$). Logit is each output obtained after the fully-connected layer as $logit_i=G(F_i(x))$ where $G$ denotes the fully-connected layer, $F_i$ denotes the feature extractor with the $i$-th normalization layer, and where $x$ denotes the input test data. On the other hand, the probability is the output after the softmax function and can be calculated as $prob_i=softmax(G(F_i(x)))$.

\paragraph{Average logit and average probability}
As the method above, two options can be used to choose the final result, which is based on the average logit ($\frac{1}{N}\sum_{i=1}^{N}logit_i$) \cite{seo2020learning, Chang_2019_CVPR} or average probability ($\frac{1}{N}\sum_{i=1}^{N}prob_i$). 

\paragraph{Selection based on the Wasserstein distance}
A normalization layer that minimizes the sum of Wasserstein distances between the source domain normalization statistics and instance-level feature statistics is selected at each layer. Specifically, the normalization statistics which had the smallest distance were used for normalization in each layer. The distance is computed as the sum of the Wasserstein distance between the mean values of the normalization layer and the input instance-level features, as well as the Wasserstein distance between the standard deviation values of the normalization layer and the input instance-level features, i.e $W(\mu_{\textbf{z}}, \mu_{Norm_i}) +W(\sigma_{\textbf{z}}, \sigma_{Norm_i})$, where $\mu_{Norm_i}$ and $\sigma_{Norm_i}$ denotes the pre-calculated average and standard deviation of $i$-th normalization, and $\mu_{\textbf{z}}$ and $\sigma_{\textbf{z}}$ denotes the mean and standard deviation of instance-level feature $\textbf{z}$ obtained from input data. The average and standard deviation are computed in the channel dimension \cite{kim2022dg}.

\paragraph{Selection based on the Euclidean distance}
As the method mentioned above, a normalization layer that minimizes the sum of Euclidean distances is selected at inference, as  $\sqrt{(\mu_{\textbf{z}}-\mu_{Norm_i})^2+(\sigma_{\textbf{z}}-\sigma_{Norm_i})^2}$.

\section{EXPERIMENTS}
\subsection{Dataset Description and Preprocessing}
We utilized an openly accessible dataset \cite{cui2022eeg, kim2023calibration}, comprising EEG signals from eleven individual subjects drawn from a dataset \cite{cao2019multi} conducted at the National Chiao Tung University in Taiwan. The EEG signals were recorded during a 90-minute driving session on an empty, straight road. Participants were tasked with maintaining their focus on the road and steering the wheel in response to random lane deviations. We categorized the data samples into two classes: `alert' and `drowsy,' based on the participants' reaction time (RT). RT refers to the time difference between the start of the lane-deviation event and the start of the driver's response \cite{ cui2022eeg}. EEG signals were recorded using 32 Ag/AgCl electrodes and re-sampled at a rate of 128 Hz.

Among the eleven selected subjects, we maintained a relatively balanced distribution of these classes across different sessions, ensuring that each class had more than 50 samples. This resulted in a total of 1,221 samples for the `drowsy' class and 1,731 samples for the `alert' class.

\subsection{Implementation Details}
We evaluated the performance using leave-one-subject-out cross-validation \cite{kim2022dg, kim2023calibration}. The detailed hyperparameters are as in \cite{kim2023calibration}. In addition, the `alert' class was accounted as a positive class while calculating the metrics.

\begin{table}[t!]
    \caption{Average drowsiness classification performance (\%) of deep neural networks}
    \centering
    \resizebox{\columnwidth}{!}{%
    \begin{tabular}{lccccc} \toprule
    Model    & Accuracy   & \textit{F}1-score & Precision & Recall & AUROC  \\ \midrule
    DeepConvNet \cite{schirrmeister2017deep} & 70.45 & 56.73 & 76.11 & 49.39 & 66.91 \\
    EEGNet4,2 \cite{lawhern2018eegnet}    & 69.32 & 55.57 & 74.50 & 49.41 & 66.47 \\
    EEGNet8,2 \cite{lawhern2018eegnet}    & 70.92 & 60.94 & 74.23 & 56.84 & 69.05 \\
    ResNet1D-8 \cite{han2021domain}  & \textbf{72.93} & \textbf{69.09} &\textbf{72.28} & \textbf{72.97} & \textbf{72.74} \\
    ResNet1D-18 \cite{kim2022dg}  & 71.42 & 67.70 & 73.68 & 70.01 & 71.93  \\ \bottomrule    
    \end{tabular}}
    \label{tab:baseline}
\end{table}

\subsection{Results and Discussion}
\subsubsection{Deep neural networks}
Table \ref{tab:baseline} shows the average drowsiness classification performance of five deep neural networks. ResNet1D-8 achieved the highest accuracy of 72.93\%, \textit{F}1-score of 69.09\%, precision of 72.28\%, recall, in other words, sensitivity, of 72.97\%, and area under the receiver operating characteristics (AUROC) of 69.33\%. 
Among EEGNet models, EEGNet8,2 achieved a higher \textit{F}1-score of 60.94\% and the highest AUROC of 69.05\%. Among ResNet1D models, the network with shallow layers achieved a higher generalization performance, an \textit{F}1-score of 64.67\% and AUROC of 69.33\%. We believe that the total number of data samples influenced the depth of the model. As a result, we selected the ResNet1D-8 as our backbone network for applying various normalization methods.

\begin{table}[t!]
    \caption{Average drowsiness classification performance (\%) according to the type of normalization}
    \centering
    \resizebox{\columnwidth}{!}{%
    \begin{tabular}{lccccc} \toprule
    Normalization & Accuracy    & \textit{F}1-score & Precision & Recall & AUROC  \\ \midrule
    BN & \textbf{72.93} & 69.09 & 72.28 & 72.97 & \textbf{72.74} \\
    IN & 47.41 & 41.27 & 34.29 & 57.83 & 46.97 \\
    IBN & 71.93 & \textbf{69.25} & \textbf{72.54} & \textbf{74.68} & 72.28 \\ \midrule
    DSBN \cite{Chang_2019_CVPR} & \textbf{73.27} & \textbf{69.63} & \textbf{71.98} & 74.17 & \textbf{74.02} \\
    DSIN & 47.40 & 55.24 & 44.39 & \textbf{78.31} & 47.87 \\
    DSON \cite{seo2020learning} & 71.08 & 65.80 & 70.64 & 70.35 & 69.56  \\\bottomrule
    \end{tabular}}
    \label{tab:compareBN}
\end{table}

\begin{table}[t!]
    \caption{Average classification performance (\%) based on the method used for inference}
    \centering
    \resizebox{\columnwidth}{!}{%
    \begin{tabular}{lcccccc} \toprule
    Method  & Accuracy &\textit{F}1-score & Precision & Recall & AUROC  \\ \midrule
    \addstackgap[3.5pt] {Max logit}  & 67.88 & 61.75 & 65.14 & 65.24 & 67.73 \\
    \addstackgap[3.5pt] {Max prob.} & 68.64 & 65.47 & 68.90 & 71.87 & 70.22 \\
    \addstackgap[3.5pt] {Average logit} & 72.63 & 68.47 & 71.58 & 72.42 & 73.34 \\
    \addstackgap[3.5pt] {Average prob.} & \textbf{73.27} & \textbf{69.63} & \textbf{71.98} & \textbf{74.17} & \textbf{74.02} \\
    \thead[l]{Select based on \\ Wasserstein dist.} & 69.75 & 63.41 & 66.27 & 65.08 & 68.42 \\
    \thead[l]{Select based on \\ Euclidean dist.} & 68.62 & 65.10 & 65.09 & 70.78 & 68.42    \\ \bottomrule    
    \multicolumn{5}{l}{prob.: Probability, dist.: Distance}\\
    \end{tabular}}
    \label{tab:compareInfer}
\end{table}

\subsubsection{Domain-invariant normalization}
Table \ref{tab:compareBN} shows the average drowsiness classification performance according to the types of normalization. Applying a domain-invariant BN resulted in the highest accuracy of 72.93\% and AUROC of 72.74\%, and the second highest \textit{F}1-score of 69.09\%, precision of 72.28\%, and recall of 72.97\%. As in previous studies, applying domain-invariant IN has decreased the discriminative performance over classes which yielded performance similar to chance-level \cite{seo2020learning}. On the other hand, utilizing a mixture of IN and BN showed the highest \textit{F}1-score, recall, and AUROC of 69.25\%, 72.54\%, and 74.68\%, respectively. The mixture of IN and BN had optimized each other's trade-offs \cite{pan2018two}.

\subsubsection{Domain-specific normalization}
As shown in Table \ref{tab:compareBN}, applying DSBN achieved the highest overall performance, an accuracy of 73.27\%, \textit{F}1-score of 69.63\%, precision of 71.98\%, and AUROC of 74.02\%. Domain-specific IN resulted in the highest recall of 78.31\% and the lowest performance in the remaining evaluation metrics. Moreover, applying domain-specific batch normalization outperformed domain-invariant normalization. 

The final results of models with domain-specific normalization are computed based on the average probability, as shown in Table \ref{tab:compareInfer}. Table \ref{tab:compareInfer} shows the average drowsiness classification performance based on the method used for inference. Among max, average, and selection methods, the average method yielded the highest performance, while the selection specifically selecting BN at each layer based on the Wasserstein distance resulted in the lowest \textit{F}1-score, recall, and AUROC of 63.41\%, 65.08\%, and 68.42\%, respectively. Additionally among logits and probabilities, using the probabilities yielded higher performance.

\begin{figure}[t!]
    \includegraphics[width=\columnwidth]{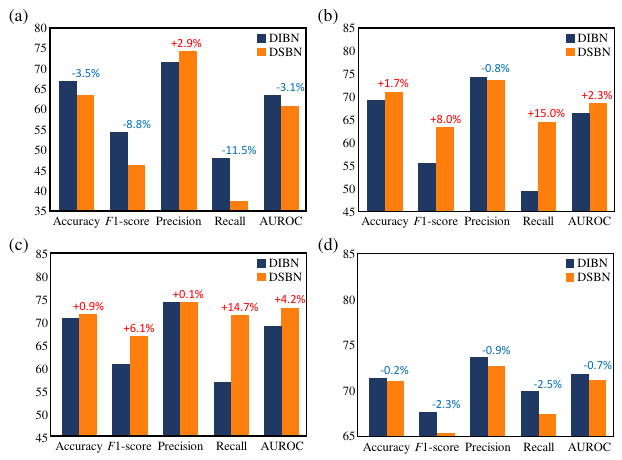}
    \centering
    \caption{Ablation study of other deep neural networks based on the presence of domain-specific batch normalization (DSBN). (a) DeepConvNet, (b) EEGNet4,2, (c) EEGNet8,2, (d)ResNet1D-18. The percentage over the bar (performance of the model with DSBN) denotes the performance difference with the model with domain-invariant batch normalization (DIBN). The final label prediction in the inference phase is determined based on the average value of probabilities.}
    \label{fig:result}
\end{figure}

\subsubsection{Ablation study}
We conducted an ablation study on the presence of DSBN in other deep neural networks. Fig. \ref{fig:result} shows the performance comparison of each deep neural network based on the presence of DSBN. The final label prediction in the inference phase was determined based on the average value of probabilities. As shown in Fig. \ref{fig:result}, DeepConvNet with DSBN generally showed a decrease in performance, and EEGNet4,2, EEGNet8,2, and ResNet1D-8 showed an overall increase in performance. Specifically, EEGNet4,2 with DSBN yielded the highest increase in performance especially in the recall metric. And among EEGNet models, EEGNet8,2 outperformed an \textit{F}1-score of 67.04\%, recall of 64.42\%, and AUROC of 68.75\%. Having more temporal filters contributed to better performance in the EEG signal dataset. Among the ResNet1D models, the performance of ResNet1D-18 showed a decrease in all metrics.

\section{CONCLUSION}
We proposed a robust framework for classifying driver drowsiness utilizing domain-specific batch normalization. Statistics of each domain were computed by separate normalization. Throughout the experiments with various normalization methods, we claim that domain-specific batch normalization, particularly utilizing the average of logits in making the final result for inference improved the generalization performance, i.e. improved the ability to generalize across subjects. We will benchmark additional datasets related to drivers' mental states and compare other normalization methods in future research. 

\bibliographystyle{IEEEtran}
\bibliography{REFERENCE}

\end{document}